\documentclass[aps,prl,twocolumn,amsmath,amssymb,showpacs,superscriptaddress,floatfix,longbibliography]{revtex4-1}
\usepackage{framed}
\usepackage{mathtools}
\usepackage{graphicx}
\usepackage{dcolumn}
\usepackage{amsmath}
\usepackage{amssymb}
\usepackage{subfigure,amsmath,verbatim,moreverb,bm}
\usepackage{color}

\def\be{\begin{equation}}
\def\ee{\end{equation}}
\def\ber{\begin{eqnarray}}
\def\eer{\end{eqnarray}}

\def\rv{{\bf r}}
\def\nv{{\bf n}}

\def\jv{{\bf j}}

\def\Sv{{\bf S}}

\begin{document}

\title{Boundary conditions for spin and charge diffusion in the presence of interfacial spin-orbit coupling}

\author{J. Borge}
\email{juanborge@hotmail.com} 
\affiliation{Nano-Bio Spectroscopy group, Departamento F\'isica de Materiales,
Universidad del Pa\'is Vasco, Av. Tolosa 72, E-20018 San Sebasti\'an, Spain} 

\author{I. V. Tokatly}
\email{ilya.tokatly@ehu.es} 
\affiliation{Nano-Bio Spectroscopy group, Departamento  F\'isica de Materiales, 
Universidad del Pa\'is Vasco, Av. Tolosa 72, E-20018 San Sebasti\'an, Spain} 
\affiliation{IKERBASQUE, Basque Foundation for Science, E48011 Bilbao, Spain} 
\affiliation{Donostia International Physics Center (DIPC),Manuel de Lardizabal 4, E-20018 San Sebastian, Spain}

\begin{abstract}
Breaking of the inversion symmetry at the interface between different materials may dramatically enhance  spin-orbit interaction in the vicinity of the interface. We incorporate the effects of this interfacial spin-orbit coupling (ISOC) into the standard drift-diffusion theory by deriving generalized boundary conditions for diffusion equations. Our theoretical scheme is based on symmetry arguments, providing a natural classification and parametrization of all spin-charge and spin-spin conversion effects that occur due to ISOC at macroscopically isotropic interfaces between nonmagnetic materials. We illustrate our approach with specific examples of spin-charge conversion in hybrid structures. In particular, for a lateral metal-insulator structure we predict an ``ISOC-gating'' effect which can be used to detect spin currents in metallic films with weak bulk SOC. 
\end{abstract}

\maketitle

Correlations between charge and spin degrees of freedom induced by spin-orbit coupling (SOC) in crystals and nanostructures open a pathway to control spin dynamics by purely electric means, without using magnetic fields. Not surprisingly, spin-charge conversion phenomena mediated by SOC are attracting a growing attention in the field of spintronics \cite{Zutic04, Sinova2015, Handbook,TenYears2010}.
Among them, the most known are the spin Hall effect (SHE) \cite{Dyakonov71,Kato04, Sih05}, and the inverse spin-galvanic effect also known as the Edelstein effect \cite{Lyanda-Geller89,Edelstein1990} (EE).
The SHE universally exists in all conductors without any symmetry restriction, provided SOC is sufficiently strong. In particular, it is responsible for the spin-charge conversion in the bulk of centrosymmetric materials, like Pt or Au \cite{Valenzuela_Nat06,Takahashi_Revese_PRL07,Takahashi_GSH_NatMater08}.
In contrast, the EE, that is, the spin polarization induced by a charge current \cite{Ganichev02,Ioan2010,Kato04,Sih05,Inoue2003,Yang2006}, occurs only in the absence of inversion symmetry or, more precisely, only in gyrotropic materials/structures \cite{GanTruSch2016}. 
Usually it is discussed for two-dimensional (2D) electron gases in semiconductor heterostructures or in surface bands at surfaces or interfaces \cite{Sih05,Norman2014,Seibold2017,Shen2014}. SOC also leads to the spin-spin conversion via the spin swapping effect (SSE) \cite{Lifshits2009,Shen04,Saidaoui2016}.

The symmetry conditions for all spin-charge conversion effects are naturally met at interfaces between different materials as any interface is always locally gyrotropic. Moreover the strong inversion symmetry breaking across the interface dramatically enhances manifestations of SOC, and, depending on the nature of the materials, may produce a giant interfacial SOC (ISOC) \cite{Ast2007,Mathias2010,Rybkin2010,Moreschini2009}. This  makes interfaces promising candidates 
for active regions in spintronics devises, where the spin-charge and spin-spin conversion occur most efficiently. In the last years these effects have been measured using different experimental techniques for various interfaces \cite{Rojas2013,Isasa2016,Karube2015,Karube2016,Miron2011,Baek2018}.

First experiments on the spin-charge conversion due to ISOC were interpreted as the inverse EE (IEE) in the 2D Rashba-splitted interface band \cite{Rojas2013,Shen2014}. Later it has been recognized that the spin-dependent scattering of the bulk continuum states at the interface also contributes strongly to the interfacial spin-charge conversion and the spin swapping \cite{Wang2013,Borge2014,Tokatly2015,Borge17,Li2018}. Currently, theoretical studies of spin transport in the presence of ISOC are limited to specific effects in specific microscopic models with simplest geometries. Apparently this is not sufficient for the description of realistic device structures, and it is highly desirable to classify all effects of ISOC and consistently incorporate them into a general theoretical scheme of device modeling. 

The spin and charge transport in a typical spintronics device is usually well described by the drift-diffusion theory. Within this approach the evolution of the spin and charge densities is governed by diffusion equations \cite{DyaKha2017}, supplemented with proper boundary conditions (BC) at all interfaces and boundaries. In the absence of SOC the BC reduce to the conservation of normal to the interface components of all currents, and relations between the currents and possible discontinuities of the densities across the interface. The latter are usually formulated in terms of spin-dependent interface conductances \cite{Hernando2000,BraNazBau2001,BraBauKel2006}. The modifications of BC by the bulk SOC in noncentrosymmetric materials have been intensively debated in the literature \cite{Malshukov2005,AdaBau2005,GalBurSar2006,Bleibaum2006,Tserkovnyak2007}. However the role of ISOC and the ways of incorporating its effects into the BC for the drift-diffusion theory remain largely unexplored. Recently a 
generalization of the magnetoelectronic circuit theory, which partly accounts for the ISOC via coupling to the in-plane electric field at the interfaces has been proposed \cite{Amin2016-1,Amin2016-2}. This indeed captures the interfacial generation of spin current by the in-plane charge current \cite{Amin18,Baek2018}, but apparently it does not cover all physically expected effects of ISOC and the general form of the corresponding BC still remains unknown.

The present paper is aimed at filling this gap by deriving the full set of BC describing all possible spin-charge and spin-spin conversion effects that may occur at the macroscopically isotropic interface separating non-magnetic materials. We do not use any specific microscopic model, but rely solely on symmetry arguments, which is similar to the symmetry based derivation of spin diffusion equations in the presence of bulk SOC \cite{Dyakonov2007,DyaKha2017}. 

Let us consider two nonmagnetic materials labeled by the index $\alpha=1,2$ and separated by a flat interface characterized by unit normal vector $\hat{\nv}$. The interface located at the surface $\hat{\bf n}\cdot\rv=0$ is assumed macroscopically isotropic with a symmetry group $C_{\infty v}$. In the bulk of the materials the charge and spin degrees of freedom are described, respectively, by the distribution of electrochemical potentials $\mu_{\alpha}({\bf r})$ and the spin density ${\bf S}_{\alpha}({\bf r})$. To focus on the effects of ISOC, we assume that both materials posses the inversion symmetry and the spin-charge coupling in the bulk is negligible. In this case the charge 
${\bf j}_{\alpha}$ and spin $J_{\alpha i}^a$ currents are given by the standard diffusive formulas, ${\bf j}_{\alpha}=-\sigma^D_{\alpha}\nabla\mu_{\alpha}$, and $J_{\alpha i}^a=-D_{\alpha}\partial_iS_{\alpha}^a$, where $\sigma^D_{\alpha}$ and $D_{\alpha}$ are the Drude conductivity and the diffusion coefficient, respectively. In the steady state the charge and spin diffusion equations on either side of the interface reduce to Laplace equations for $\mu_{\alpha}({\bf r})$, and the stationary spin diffusion equations,
\begin{equation}
\label{L1}
\nabla^2\mu_{\alpha}({\bf r})=0\; ; \qquad
D_{\alpha}\nabla^2{\bf S}_{\alpha}({\bf r})=\frac{{\bf S}_{\alpha}({\bf r})}{\tau_{\alpha}},
\end{equation}
where $\tau_{\alpha}$ is the spin relaxation time.
In the absence of ISOC the BC at the interface are well known and read
\begin{eqnarray}
\label{BC-charge-0}
&&\sigma^D_\alpha({\bf \hat{n}}\cdot \nabla)\mu_\alpha=G_0^c\Delta\mu, \\
 \label{BC-spin-0}
&&D_\alpha({\bf \hat{n}}\cdot \nabla){\bf S}_\alpha=G_0^s\Delta{\bf S},
\end{eqnarray}
where $\Delta\mu=\mu_1-\mu_2$ and $\Delta{\bf S}={\bf S}_1-{\bf S}_2$, and $G_0^{c/s}$ is the charge/spin conductance \cite{BCnotation-note}. 
Physically Eqs.~\eqref{BC-charge-0} and \eqref{BC-spin-0} relate the currents 
passing through the interface to the interfacial density/potential drops. The appearance of the differences of the densities in the BC, and the independence of conductances on the material index $\alpha$ reflect the conservation of all currents in the absence of SOC.

Formally Eqs.~\eqref{BC-charge-0}-\eqref{BC-spin-0} are linear relations between the densities and their first derivatives. Such relations are forbidden by the symmetry in the isotropic bulk, but they are allowed at the interface as it provides us with an additional polar vector $\hat{\bf n}$. By constructing a scalar differential operator $\hat{\bf n}\cdot\nabla$ we can compile linear relations involving the densities and their derivatives, and transforming as a scalar,  Eq.~\eqref{BC-charge-0}, and a pseudovector, Eq.~\eqref{BC-spin-0}. These are the general BC for the scalar $\mu(\rv)$ and the pseudovector $\Sv(\rv)$ densities, allowed by the interface $C_{\infty v}$ symmetry under the requirements of the charge and spin conservation in the absence of the charge-spin mixing. 

In the presence of ISOC spin is not conserved and only the charge conservation (the gauge invariance) requirement remains. This allows for additional terms in the BC. Let us consider first the modification of the scalar BC in Eq.~\eqref{BC-spin-0}. The only additional scalar invariant that is linear in the densities and their first derivatives is $({\bf \hat{n}}\times\nabla)\cdot{\bf S}$. Therefore the most general scalar BC takes the form,
\begin{equation}
\label{BC-charge}
\sigma^D_\alpha({\bf \hat{n}}\cdot \nabla)\mu_\alpha = G\Delta\mu +  
 \sum_{\beta}\theta^{sc}_{\alpha\beta}D_{\beta}({\bf \hat{n}}\times\nabla)\cdot {\bf S}_{\beta}.
\end{equation}
Because of the gauge invariance the electrochemical potentials enter only as $\Delta\mu$, and there is only one charge conductance $G$. The second term in Eq.~\eqref{BC-charge} describes the spin-charge conversion via the interfacial ISHE -- the generation of a normal charge current from in-plane spin currents at either side of the interface. This channel of the spin-charge conversion at hybrid interfaces has been discussed in Ref.~\onlinecite{Linder2011} within a simple ballistic scattering model. Our symmetry arguments show that in general it is parametrized by four spin-charge Hall angles $\theta^{sc}_{\alpha\beta}$. The cross-interface angles $\theta^{sc}_{12}$ and $\theta^{sc}_{21}$ should vanish for nontransparent interfaces. For example, for metal-insulator interfaces there is by only one spin-charge Hall angle.    

Similarly we generalize the pseudovector BC of Eq.~\eqref{BC-spin-0} by adding all symmetry allowed pseudovectors constructed from the densities and their derivatives \cite{SM-note}. It is convenient to write the resulting general BC by separating the normal and the parallel to the interface spin components ${\bf S}={\bf S}_{\perp}+{\bf S}_{\parallel}$, where 
${\bf S}_{\perp}={\bf \hat{n}}({\bf \hat{n}}\cdot{\bf S})$ and
${\bf S}_{\parallel}=({\bf \hat{n}}\times{\bf S})\times{\bf \hat{n}}$,
\begin{widetext}
\begin{eqnarray}
\label{BC-spin-n}
&&D_{\alpha}({\bf \hat{n}}\cdot \nabla){\bf S}_{\alpha \perp} = 
G^{n}_{\alpha}\Delta{\bf S}_{\perp} + L^{n}_{\alpha}\bar{\bf S}_{\perp} + 
\sum_{\beta}\kappa^{n}_{\alpha\beta}D_{\beta}({\bf\hat{n}}\times\nabla)\times{\bf S}_{\beta \parallel}\\
\label{BC-spin-p}
&&D_{\alpha}({\bf \hat{n}}\cdot \nabla){\bf S}_{\alpha \parallel} =
G^{p}_{\alpha}\Delta{\bf S}_{\parallel} + L^{p}_{\alpha}\bar{\bf S}_{\parallel} + 
\sum_{\beta}\kappa^{p}_{\alpha\beta}D_{\beta}({\bf\hat{n}}\times\nabla)\times{\bf S}_{\beta \perp} +
\sum_{\beta}\theta^{cs}_{\alpha\beta}\sigma^D_{\beta}({\bf \hat{n}}\times\nabla)\mu_{\beta},
\end{eqnarray}
\end{widetext}
where $\bar{\bf S}=\Sv_1+\Sv_2$. In the presence of ISOC the spin is not conserved. Therefore the right and left values of the boundary spin can independently enter BC. The corresponding contributions are parametrized by the spin conductances $G^{n/p}_{\alpha}$ and the spin loss coefficients $L^{n/p}_{\alpha}$, which in general depend on the material index $\alpha$, and are different for the normal ($n$) and the parallel ($p$) spin components. 
The third term in the right hand sides in Eqs.~\eqref{BC-spin-n} and \eqref{BC-spin-p} describes the spin-spin conversion due to the interfacial SSE. Namely, the in-plane current of the parallel (normal) spin component generates the normal current of the normal (parallel) spin component. This effect of ISOC is characterized by a set of swapping coefficients $\kappa^{n/p}_{\alpha\beta}$. Finally, the last term in the right hand side in Eq.~\eqref{BC-spin-p} is responsible for the charge-spin coupling. It can be interpreted as an interfacial SHE -- generation of the spin current across the interface by an in-plane charge current. This effect has been studied recently in Refs.~\onlinecite{Amin18,Baek2018} for different hybrid structures via first principle transport calculations. The corresponding transport coefficients in Eq.~\eqref{BC-spin-p} are the charge-spin Hall angles $\theta^{sc}_{\alpha\beta}$.

Equations \eqref{BC-charge}-\eqref{BC-spin-p} generalize the standard BC of Eqs.~\eqref{BC-charge} and \eqref{BC-spin-0}. However this is not sufficient to fully describe the physics of interfaces with ISOC. 
The reason is that the diffusion equations and the derived BC involve only smooth ``diffusive'' parts of the densities that vary slowly on the scale of the mean free path $\ell$. In addition, strongly localized (on the scale less than $\ell$) 
interfacial charge and spin currents as well as the interfacial spin polarization in general appear near nontrivial spin-orbit active interfaces. Most obviously the localized observables can be related to the interface bands, as it is commonly assumed to interpret experiments on the interfacial spin-charge conversion \cite{Rojas2013,Karube2016,Shen2014,Seibold2017}. Apart from that, in the presence of ISOC the spin-dependent interference between the incident and reflected waves for bulk states also leads to the appearance of interfacial spin polarization \cite{Tokatly2015} and interfacial currents \cite{Borge17}, localized on the scale of the Fermi wavelength $\lambda_F$. 

The localized contributions can be included into the drift-diffusion theory by representing the total physical observables in the following form,
\begin{equation}
\label{I-observ-def}
O_{tot}({\bf r})=\Theta(-z)O_1({\bf r}) +\Theta(z)O_2({\bf r}) + \delta(z)O_I({\bf r}_{\parallel}),
\end{equation}
where $z=\hat{\nv}\cdot\rv$ is the normal to the interface coordinate, $O_{\alpha}(\rv)$ are the slow ``diffusive'' parts that satisfy the bulk diffusion equations, and $O_I({\bf r}_{\parallel})$ is the localized part of the observable. Within the standard linear transport theory the localized spin $\Sv_I$, the charge current $\jv_I$, and the spin current $J_{iI}^a$ should be related linearly to the interfacial values of the diffusive observables $\mu_{\alpha}$ and $\Sv_{\alpha}$. Formally the latter act as the sources (effective driving fields) for the former. The general form of such relations for $\Sv_I$, $\jv_I$, and $J_{iI}^a$ can be determined from the symmetry arguments by combining, respectively, all linearly independent pseudovector, vector, and pseudotensor invariants constructed out of $\mu_{\alpha}$, $\Sv_{\alpha}$ and their first derivatives \cite{n-derivative-note}. 
A straightforward analysis \cite{SM-note} leads to the following expressions for the localized parts of the spin polarization and the charge current,
\begin{eqnarray}
\label{S-I}
{\bf S}_I&=&\sum_{\alpha}\sigma^{cs}_{\alpha}({\bf \hat{n}}\times\nabla )\mu_{\alpha} + 
\sum_{\alpha}\sigma^{ss}_{\alpha}({\bf \hat{n}}\times\nabla)\times {\bf S}_{\alpha}\\
\label{j-I}
{\bf j}_{I}&=&\sum_{\alpha} \sigma^{sc}_{\alpha}({\bf \hat{n}}\times{\bf S}_{\alpha}) +
\sum_{\alpha} \theta^{sc}_{I\alpha}({\bf \hat{n}}\times\nabla )({\bf \hat{n}}\cdot {\bf S}_{\alpha}),
\end{eqnarray}
while the localized spin current takes the form,
\begin{widetext}
\begin{equation}
\label{J-I}
J_{iI}^a = g^{cs}\epsilon_{iak}\hat{n}_k\Delta\mu 
+ \sum_{\alpha}\Big[g^{p}_{\alpha}\hat{n}_aS^i_{\alpha}
+ g^{n}_{\alpha}\delta_{ai}\;{\bf n\cdot S_{\alpha}} 
+  \theta^{cs}_{I\alpha}\hat{n}_a({\bf \hat{n}}\times\nabla)_i\mu_{\alpha}
+ \kappa_{I\alpha} ({\bf \hat{n}}\times({\bf \hat{n}}\times\nabla))_a S_{\alpha}^i
+ \kappa_{I\alpha}'\delta_{ai}\nabla\cdot {\bf S_{\alpha\parallel}}\Big]
\end{equation}

\end{widetext}
In the last equation the spatial index $i$ takes only in-plane values as the interface spin current $J_{iI}^a$ flows in the interface plane \cite{projector-note}.
For brevity we do not show the ``trivial'' terms proportional to $\Sv_{\alpha}$, $\nabla\mu_{\alpha}$, and $\partial_i S_{\alpha}^a$, allowed in Eqs.~\eqref{S-I}, \eqref{j-I}, and \eqref{J-I}, respectively. These terms may describe, if necessary, the usual 2D diffisive transport in the interface bands. The contributions shown explicitly are those responsible for the spin-charge and the spin-spin conversion. The first term in Eq.~\eqref{S-I} describes the interfacial EE -- the local spin polarization induced by the in-plane charge current \cite{Tokatly2015}. The second term is the interfacial spin generated by the spin current flowing along the interface, and polarized in the direction orthogonal to that of the current. The first and the second terms in Eq.~\eqref{j-I} correspond, respectively, to the interfacial IEE and the ISHE, i.~e., the charge current at the interface induced by the non-equilibrium spin polarization and the in-plane spin current. Finally, in Eq.~\eqref{J-I} the first term can be 
interpreted as a cross-interface 
SHE (the spin current at the interface plane generated by the voltage drop across the interface), the fourth term is the 2D interfacial SHE, the last two terms describe the 2D SSE, while the second and third terms are responsible for the spin current produced directly by the non-equilibrium spin polarization.

The localized observables of Eqs.~\eqref{S-I}-\eqref{J-I} were introduced on physical grounds. Now we show that the appearance of in-plane localized currents is also required by the internal consistency of the theory. Let us look on Eq.~\eqref{BC-charge}. The second term in the right hand side is allowed by the symmetry and meaningful physically, but it manifestly violates conservation of the charge current. Indeed Eq.~\eqref{BC-charge} states that a part of the charge current passing through the interface is lost in the presence of in-plane spin gradients. The interfacial charge current $\jv_I$ fixes this problem by providing a missing sink. In the presence of $\jv_I$ the continuity equation for the total charge current, after the integration across the interface, reads,
\begin{equation}\label{I-continuity}
 \sigma^D_1(\hat{\nv}\cdot\nabla)\mu_1-\sigma^D_2(\hat{\nv}\cdot\nabla)\mu_2=-\nabla\cdot\jv_I.
\end{equation}
By substituting Eqs.~\eqref{BC-charge} and \eqref{j-I} into the left and right hand sides we find that the charge continuity equation is fulfilled identically if the spin-charge Hall angles $\theta^{sc}_{\alpha\beta}$ are related to the ``Edelstein conductivity'' $\sigma^{sc}_{\alpha}$ as follows,
\begin{equation}
 \label{theta-sigma}
\sigma^{sc}_{\alpha} =D_{\alpha}(\theta^{sc}_{1\alpha} - \theta^{sc}_{2\alpha}).
\end{equation}
Therefore there is a deep connection between the inverse SHE described by Eq.~\eqref{BC-charge} and the generation of the local charge current via the inverse EE in Eq.~\eqref{j-I}. Inclusion of one effect necessarily implies the presence of the other.

The BC Eqs.~\eqref{BC-charge}-\eqref{BC-spin-p} together with Eqs.~\eqref{S-I}-\eqref{J-I} complement the standard bulk drift-diffusion equations to model spintronics devices of any experimentally relevant geometry. It is worth noting that technically the localized currents become important in non-1D geometries with interfaces of a finite size. At the edges of the interface the total currents should be conserved, and therefore the edges act as local sources and sinks, which generates nontrival patterns of the charge and spin flows. The examples below illustrate this point and demonstrate our general phenomenological construction at work.

\begin{figure}
\begin{center}
\includegraphics[width=3.2in]{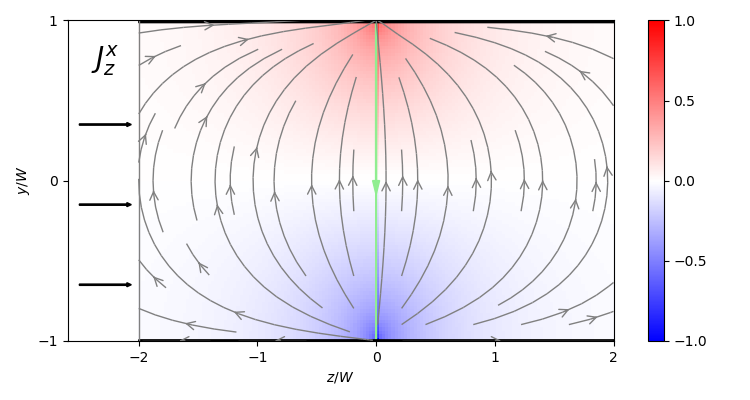}
\caption{Streamlines of the charge current generated by the spin current $J_z^x$ flowing perpendicular to the interface (at $z=0$) between two conducting materials. The density plot shows the distribution of the induced electrostatic potential.}
\label{Currents-1}
\end{center}
\end{figure}

The first example models the spin-charge conversion at the interface \cite{Rojas2013}. We consider a conducting bilayer of a finite width $W$ in the $y$-direction and separated by the interface with ISOC at $z=0$ plane, as shown in Fig.~\ref{Currents-1}. 
A spin current $J_z^x(z)=-D\partial_zS^x(z)$, polarized along $x$-axis, and injected from the left, flows in $z$-direction, crosses the interface, and determines via Eq.~\eqref{BC-spin-p} the spin polarization $S^x(0)$ at the interface. The latter, in turn, generates a localized charge current in the y-direction via the IEE in Eq.~\eqref{j-I}, $j_{yI}=\sigma^{sc}S^x(0)$. To determine the distribution of the potential $\mu(\rv)$ and the charge current $\jv=-\sigma^D\nabla\mu$ in the bulk we have to solve the Laplace equation, $\nabla^2\mu=0$, with the condition of vanishing normal component of the {\em total} current at the sample boundaries at $y=\pm W/2$ 
\begin{equation}
\label{BCJ}
-\sigma^D\partial_y\mu(y,z)|_{y=\pm W/2}+j_{yI}\delta(z)=0.
\end{equation}
By solving this problem analytically \cite{SM-note} we find the spatial distribution of the charge current,
and the total voltage drop across the sample
\begin{equation}
\Delta V=\int\big[\mu(W/2,z)-\mu(-W/2,z)\big]dz=j_{yI}\frac{W}{\sigma^D},
\end{equation}
The stream lines of the induced charge flow together with the density plot for the potential are shown in Fig.~\ref{Currents-1}. The current in the bulk forms a counterflow that compensates the localized currents generated at the interface. Both the induced potential and the current are concentrated near the edges of the interface at a macroscopic scale of the order of the sample size $W$.

\begin{figure}
\begin{center}
\includegraphics[width=3.2in]{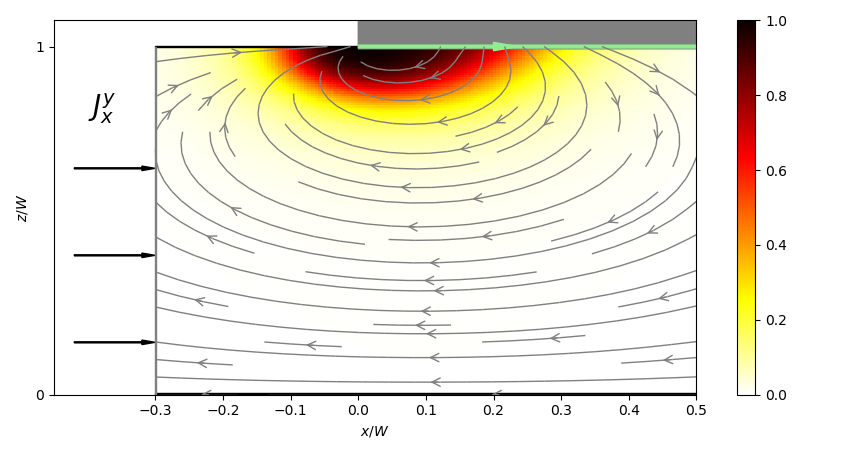}
\caption{Charge flow generated by the spin current $J_x^y(x)$ in the metallic film with a insulator (shown in gray) deposited on its top surface. The density plot shows the current strength.}
\label{Currents-2}
\end{center}
\end{figure}

As a second example we consider a spin-charge conversion in a lateral hybrid structure made from a metallic film of thickness $W$, with a part of its upper surface covered by an insulator with large SOC, like Bi$_2$O$_3$, see Fig.~\ref{Currents-2}. 
In this way we create an interface with ISOC on the top boundary at $z=W$ for $x>0$, while the rest ($x<0$) of the top boundary as well as the bottom boundary at $z=0$ remain ``trivial''. We assume a diffusive spin current polarized along $y$, flowing in the $x$-direction, that is, $J_x^y(x)=-D\partial_xS^y(x)$ with 
$S^y(x)\sim e^{-x/l_s}$, where $l_s=\sqrt{D\tau_s}$ is the spin diffusion length. 

The induced potential $\mu(x,z)$, is obtained by solving the Laplace equation with two BC. On the bottom surface the standard BC of $\sigma^D\partial_z\mu|_{z=0}=0$ is imposed. To get the BC on the top surface we combine Eqs.~\eqref{BC-charge} and \eqref{j-I} in form of Eq.~\eqref{I-continuity} that for the metal-insulator interface and the chosen spin density reads, 
\begin{equation}\label{BC-lateral}
  \sigma^D\partial_z\mu|_{z=W} = -D\partial_x\big[\theta^{sc}(x)S^y(x)\big],
\end{equation}
were $\theta^{cs}(x)=\theta^{sc}\Theta(x)$ reflects the stepwise distribution of ISOC at the top surface  \cite{sigma-theta-note}. This problem is also solvable analytically \cite{SM-note}. The corresponding charge flow, shown in Fig.~\ref{Currents-2}, demonstrates a typical dipolar pattern with a local sink at the edge of the interface a distributed $\sim\Theta(x)e^{-x/l_s}$ source. As the film has a finite width this dipole filed generates a lateral voltage drop: 
\begin{equation}\label{V-lateral}
 \Delta V=\mu(\infty,z)-\mu(-\infty,z)=\chi S^y(0)\theta^{sc}{l_s}/{W},
\end{equation}
where $\chi=D/\sigma^D$ is the inverse compressibility of the metal. If the top ``ISOC gate'' has a finite length $L$ the voltage drop acquires an additional factor $1-e^{-L/l_s}$. Notice that by measuring the induced lateral voltage, and using Eq.~\eqref{V-lateral} we get a direct experimental access to the interfacial spin-charge Hall angle $\theta^{sc}$.

In conclusion, we derived a full set of additional conditions that complement the standard drift-diffusion theory to model spin and charge dynamics in the presence of interfaces with strong ISOC. These conditions consist of the generalized BC describing the interfacial spin-charge and spin-spin conversion, and the expressions for the spin, the charge current, and the spin current, localized at the interface within a microscopic scale (smaller than $\ell$). Our construction provides a natural classification and parametrization of all spin-charge and spin-spin conversion effects mediated by ISOC at macroscopically isotropic interfaces between nonmagnetic materials. The phenomenological coefficients entering the derived BC should be determined from comparison with experiments or first principle calculations for specially chosen geometries. To demonstrate the working power of our theory we considered two specific examples. In particular, we predict a generation of a lateral voltage drop in a metallic film by a 
spin current if an insulator with a strong SOC is deposited on the top surface of the film. This ``ISOC gate'' effect can be used to detect spin currents in materials with weak bulk SOC.

{\it Acknowledgments}. We acknowledge financial support from the European Research Council (ERC-2015-AdG-694097),
Spanish  grant  (FIS2016-79464-P),  Grupos  Consolidados  (IT578-13),  AFOSR  Grant  No.
FA2386-15-1-0006 AOARD 144088, H2020-NMP-2014 project
MOSTOPHOS (GA No.   646259) and COST Action MP1306 (EUSpec). J.B. acknowledges  funding from
the  European Union’s Horizon 2020 research and innovation programme under the Marie Sklodowska-Curie
Grant Agreement No. 703195. J.B. acknowledges Prof. Roberto Raimondi for stimulating conversations.

\bibliographystyle{apsrev4-1}
%

\pagebreak
\onecolumngrid
\begin{center}
\textbf{\large Supplementary Material for\\ ``Boundary conditions for spin and charge diffusion in the presence of interfacial spin-orbit coupling''}
\end{center}

\setcounter{equation}{0}
\setcounter{figure}{0}
\setcounter{table}{0}
\setcounter{page}{1}
\makeatletter
\renewcommand{\theequation}{S\arabic{equation}}
\renewcommand{\thefigure}{S\arabic{figure}}
\renewcommand{\bibnumfmt}[1]{[S#1]}
\renewcommand{\citenumfont}[1]{S#1}

\section{Construction of interface tensor invariants}

In this Section we construct and classify  all possible scalars, vectors, pseudovectors and second rank pseudotensors involved in the boundary conditions (BC) of Eqs.~(4)-(6) and in the interfacial observables of Eqs.~(8)-(10) described in the main text.

\subsection{Boundary conditions}

As it is explained in the main text of the manuscript, the BC can be viewed as linear relations between the interface values of the electrochemical potentials $\mu_\alpha$, the components of the spin density $S_\alpha^a$, and their first derivatives, $\partial_i\mu_\alpha$ and $\partial_i S_\alpha^a$. In the bulk the charge and spin dynamics are described by one scalar (charge diffusion) and one pseudovector (spin diffusion) second order differential equations. Therefore we need two scalar and two pseudovector BC at each interface. Formally the problem of finding a general form of BC reduces to finding a general linear function of scalar $\mu$, pseudovector $S^a$, vector $\partial_i\mu$, and pseudotensor $\partial_i S_\alpha^a$ variables, which (i) transforms as a scalar or a pseudovector, (ii) is gauge invariant (i.~e. invariant under a global shift of the electrochemical potential), and (iii) compatible with the interface symmetry.  

The gauge invariance implies that the electrochemical potentials themselves can enter only as difference $\Delta\mu=\mu_1-\mu_2$. The compatibility with the interface symmetry formally means that the tensor coefficients in the above linear function should be invariant under operations of the interface symmetry group $C_{\infty v}$. Technically these invariant tensor coefficients can be constructed from all possible products/contractions of the vector normal to the interface $n_k$, the second-rank tensor $\delta_{ij}$, and the third-rank pseudotensor $\epsilon_{ijk}$, which are the ``natural'' available objects preserving the required symmetry.

Let us start with the scalar BC. It obviously involves the genuine scalar $\Delta\mu$. We can also generate a scalar by contracting the vector $\partial_j\mu$ with available $C_{\infty v}$ invariant vector $A_j=A n_j$, and by contracting the second rank pseudotensor $\partial_jS^b$ with the invariant second rank pseudotensor $\alpha_j^b=\alpha\epsilon_{jbk}n_k$. However, there is no way to construct a pseudovector from the available objects ($n_k$,  $\delta_{ij}$, $\epsilon_{ijk}$) in order to contract it with the pseudovector $S^b$. Hence the symmetry does not allow to make a scalar out of the spin density. To summarize, all possible scalar invariants are
\begin{equation}
 \Delta\mu;\hspace{0.3cm}A_j \partial_j\mu;\hspace{0.3cm}\alpha^{b}_j\partial_jS^b,
\end{equation}
where the symmetry allowed tensor coefficients, $A_j$ and $\alpha_j^b$, read
\begin{eqnarray}
A_j&=&A n_j\\
\alpha^{b}_j&=&\alpha\epsilon_{jbk}n_k
\end{eqnarray}
(here and in the following we use latin letters for tensors and greek letters for pseudotensors).
Taking this into account and using the vector notations we can write the most general scalar BC as follows, 
\begin{equation}
\label{DC}
 \sigma_{\alpha}^D({\bf \hat{n}}\cdot\nabla)\mu_{\alpha}=G\Delta\mu+\sum_{\beta}\theta_{\alpha\beta}^{sc}D_{\beta}({\bf \hat{n}}\times\nabla)\cdot{\bf S}_{\beta}.
\end{equation}
The indexes $\alpha$, $\beta$ indicate the side of the interface, and thus all together we have two scalar BC, as required. This equation corresponds to the Eq.~(4) of the main text.

The pseudovector BC involves the spin pseudovector $S^a$ along with other pseudovector invariants that we describe below. (i) We can make a pseudovector by contracting the vector $\partial_j\mu$ a second rank pseudotensor $\beta_j^a$. (ii)  
We may also contract a second rank tensor $B^{ab}$  with the pseudovector $S^b$. (iii) The contraction of a third rank tensor $C_j^{ab}$ with the second rank pseudotensor $\partial_jS^b$ will also generate a pseudovector. 
On the other hand, there is no way to construct a pseudovector in order to connect it with the scalar $\Delta\mu$. All the possible pseudovectors are
\begin{equation}\label{pseudovector}
 S^a;\hspace{0.3cm}\beta^{a}_j\partial_j\mu;\hspace{0.3cm}B^{ab}S^b;\hspace{0.3cm}C^{ab}_j\partial_j S^b
\end{equation}
where the tensor coefficients invariant under the operations of $C_{\infty v}$ take the following general forms 
\begin{eqnarray}\label{pseudovector-1}
\beta^{a}_{j}&=&\beta\epsilon_{jak}n_k \hspace{0.5cm}\\
\label{pseudovector-2}
B^{ab}&=&B_1(\delta_{ab}-n_an_b)+B_2n_an_b\hspace{0.5cm}\\
\label{pseudovector-3}
C^{ab}_j&=&C_1n_jn_an_b+C_2n_j(\delta_{ab}-n_an_b)+C_3n_a(\delta_{jb}-n_jn_b)+C_4n_b(\delta_{ja}-n_jn_a).
\end{eqnarray}
Notice that two terms in Eq.~\eqref{pseudovector-2} correspond to the in-plane and out-of-plane projectors. As these terms can enter with different coefficients, $B_1$ and $B_2$, the spin densities ${\bf S}_{\alpha \perp}={\bf \hat{n}}({\bf \hat{n}}\cdot{\bf S}_{\alpha})$ and ${\bf S}_{\alpha ||}=({\bf \hat{n}}\times{\bf S}_{\alpha})\times{\bf \hat{n}}$ may have different couplings. Because of this it is natural to split the pseudovector BC into two sets of equations for the normal and parallel to the interface components of the spin,
\begin{eqnarray}
\label{DS}
D_{\alpha}({\bf \hat{n}}\cdot\nabla){\bf S}_{\alpha\perp}&=&G_{\alpha}^n\Delta{\bf S}_{\perp}+L_{\alpha}^n\overline{{\bf S}}_{\perp}+\sum_{\beta}\kappa_{\alpha\beta}^nD_{\beta}({\bf \hat{n}}\times\nabla)\times{\bf S}_{\beta||}\\
\label{DS1}
D_{\alpha}({\bf \hat{n}}\cdot\nabla){\bf S}_{\alpha||}&=&G_{\alpha}^p\Delta{\bf S}_{||}+L_{\alpha}^p\overline{{\bf S}}_{||}+\sum_{\beta}\kappa_{\alpha\beta}^pD_{\beta}({\bf \hat{n}}\times\nabla)\times{\bf S}_{\beta\perp}
+\sum_{\beta}\theta_{\alpha\beta}^{cs}\sigma_{\beta}^D({\bf \hat{n}}\times\nabla)\mu_{\beta},
\end{eqnarray}
which corresponds to Eqs.~(5) and (6) of the main text.

\subsection{Localized interface observables}

Now we describe the construction of the interfacial observables introduced in the main paper. There are three possible physical objects which can be generated at the interface: the interfacial spin density, the interfacial charge current, and the interfacial spin current. The general form of linear relations of these localized observables to the ``diffusive'' variables $\mu_\alpha$,  $S_\alpha^a$, and their first derivatives, $\partial_i\mu_\alpha$ and $\partial_i S_\alpha^a$, can be determined from the symmetry arguments.

The interfacial spin density $S_I^a$ is a pseudovector and all possible pseudovector invariant have been already described above in Eqs.~\eqref{pseudovector}-\eqref{pseudovector-3}.

The interfacial charge current $j_{Ii}$ is a vector and therefore we need to construct all allowed vector invariants. A vector can be generated by (i) multiplying the scalar $\Delta\mu$ with a vector $D_i$, (ii) contracting a second rank tensor $E_{ij}$ with the vector $\partial_j\mu$, (iii) contracting a second rank pseudotensor $\gamma_i^b$ with the pseudovector $S^b$, and (iv) contracting a third rank pseudotensor $\eta_{ij}^b$ with the second rank pseudotensor $\partial_jS^b$:
\begin{equation} \label{vector}
D_i\Delta\mu;\hspace{0.3cm}E_{ij}\partial_j\mu;\hspace{0.3cm}\gamma^{b}_{i}S^b;\hspace{0.3cm}\eta_{ij}^{b}\partial_jS^b,
\end{equation}
where the allowed tensor coefficient read
\begin{eqnarray}\label{vector-1}
 D_{i}&=&D n_i \\
\label{vector-2}
 E_{ij}&=&E_1(\delta_{ij}-n_in_j)+E_2n_in_j\\
\label{vector-3}
 \gamma^{b}_i&=&\gamma\epsilon_{ibk}n_k\\
\label{vector-4}
\eta_{ij}^{b}&=&\eta_1\epsilon_{ijk}n_kn_b+\eta_2\epsilon_{jbk}n_kn_i+\eta_3\epsilon_{ibk}n_kn_j.  
\end{eqnarray}

To find the form of the interfacial spin current $J_{Ii}^a$, which is a second rank pseudotensor, we generate all allowed second rank pseudotensors. Specifically we can construct a second rank pseudotensor (i) from the scalar $\Delta\mu$ through a second rank pseudotensor $\zeta^a_i$, (ii) by contracting a third rank pseudotensor, $\xi_{ij}^a$, with the vector $\partial_j\mu$, (iii) by contracting a third rank tensor $F_i^{ab}$ with the pseudovector $S^b$, and (iv) by contracting
a fours rank tensor $G_{ij}^{ab}$ with the second rank pseudotensor $\partial_jS^b$:
\begin{equation}\label{pseudotensor}
\zeta^{a}_i\Delta\mu;\hspace{0.3cm}\xi_{ij}^a\partial_j\mu;\hspace{0.3cm}F_i^{ab}S^b;\hspace{0.3cm}G_{ij}^{ab}\partial_jS^b
\end{equation} 
where the allowed tensor coefficients have the following general form,
\begin{eqnarray}\label{pseudotensor-1}
\zeta^{a}_i&=&\zeta\epsilon_{iak}n_k\\
\label{pseudotensor-2}
\xi_{ij}^a&=&\xi_1\epsilon_{ijk}n_kn_a +\xi_2\epsilon_{jak}n_kn_i+\xi_3\epsilon_{iak}n_kn_j\\
\label{pseudotensor-3}
F_i^{ab}&=&F_1n_in_an_b+F_2n_i(\delta_{ab}-n_an_b)+F_3n_a(\delta_{ib}-n_in_b)+F_4n_b(\delta_{ia}-n_in_a)\\
\label{pseudotensor-4}
G_{ij}^{ab}&=&G_1n_in_jn_an_b+G_2n_in_j(\delta_{ab}-n_an_b)+G_3(\delta_{ij}-n_in_j)n_an_b+G_4(\delta_{ij}-n_in_j)(\delta_{ab}-n_an_b)\\
\nonumber
&+&G_5n_in_b(\delta_{ja}-n_jn_a)+G_6(\delta_{ib}-n_in_b)n_jn_a+G_7(\delta_{ib}-n_in_b)(\delta_{ja}-n_jn_a) \\ 
\nonumber
&+&G_8n_in_a(\delta_{jb}-n_jn_b)+G_9(\delta_{ia}-n_in_a)n_jn_b+G_{10}(\delta_{ia}-n_in_a)(\delta_{jb}-n_jn_b).
\end{eqnarray}
One can verify that the (pseudo)tensors appearing in each of the above equations with different coefficients form a complete set of linearly independent (pseudo)tensors of a given rank and compatible with the required symmetry.
For example, we may also construct fours rank tensors by contracting two different third rank  pseudotensors or combining them with two vectors. However, these fours rank tensors can be written as a linear combination of the terms present in $G_{ij}^{ab}$ of Eq.~\eqref{pseudotensor-4}:
\begin{eqnarray}
\epsilon_{ijl}\epsilon_{abl}&=&\delta_{ia}\delta_{jb}-\delta_{ib}\delta_{ja}\\
 \epsilon_{ijl}\epsilon_{abk}n_ln_k&=&\delta_{ia}\delta_{jb}-\delta_{ib}\delta_{ja}-\delta_{ia}n_jn_b-n_in_a\delta_{jb}+\delta_{ib}n_jn_a+n_in_b\delta_{ja}.
\end{eqnarray}

Before writing the interfacial observables in a more compact way we should take into account two facts. Firstly, the index $i$ in the current observables indicates the direction of the localized interfacial charge and spin currents. Since these currents may flow only in the interface plane we should exclude (project out) all the invariants implying the $i$-direction to be parallel to $\hat{\bf n}$. Secondly, the index $j$ indicates the direction of the diffusive charge and spin currents (the direction of the spatial derivatives of the diffusive observables). The terms corresponding to the $j$-direction parallel to $\hat{\bf n}$ should also be excluded as the normal derivatives $(\nabla\cdot{\bf \hat{n}})\mu_\alpha$ and 
$(\nabla\cdot{\bf \hat{n}}){\bf S}_\alpha$ can always be eliminated using the BC of Eqs.~(\ref{DC}, \ref{DS}, \ref{DS1}). In particular this implies that from 10 terms in Eq.~\eqref{pseudotensor-4} only 4 can appear in the expression for interface spin current, namely those with the coefficients $G_3$, $G_4$, $G_7$, and $G_8$. This leads to 4 possible contributions to the spin-spin conversion in Eq.~\eqref{J-I1} below. 

Taking into account all above arguments we can finally write the following expressions for the interface observables:
\begin{eqnarray}\label{Sp}
{\bf S}_{I||}&=&\sum_{\alpha}\chi_{I,\alpha}^{||}{\bf S}_{\alpha||} + \sum_{\alpha}\sigma^{cs}_{\alpha}({\bf \hat{n}}\times\nabla ) \mu_{\alpha} + \sum_{\alpha}\sigma^{ss}_{\alpha}({\bf \hat{n}}\times\nabla)\times {\bf S}_{\alpha\perp}\\
{\bf S}_{I\perp}&=&\sum_{\alpha}\chi_{I,\alpha}^{\perp}{\bf S}_{\alpha\perp} + \sum_{\alpha} \sigma^{ss'}_{\alpha}({\bf \hat{n}}\times\nabla)\times{\bf S}_{\alpha||}\\
\label{J1}
 {\bf j}_{I}&=&\sum_{\alpha}\sigma_{I,\alpha}^{D}{\bf \hat{n}}\times({\bf \hat{n}}\times\nabla)\mu_\alpha+\sum_{\alpha} \sigma^{sc}_{\alpha}({\bf \hat{n}}\times{\bf S}_{\alpha}) + \sum_{\alpha}\theta^{sc}_{I\alpha} ({\bf \hat{n}}\times\nabla )({\bf \hat{n}}\cdot {\bf S}_{\alpha}),
\end{eqnarray}
 and
\begin{eqnarray}
\nonumber
J_{iI}^a &=& \sum_{\alpha}\Big[D_{I,\alpha}^{||}\partial_iS^a_{||\alpha} + 
D_{I,\alpha}^{\perp}\partial_iS^a_{\perp\alpha}\Big]
+ g^{cs}\epsilon_{iak}\hat{n}_k\Delta\mu \\
&+& \sum_{\alpha}\Big[g^{p}_{\alpha}\hat{n}_aS^i_{\alpha}
+ g^{n}_{\alpha}\delta_{ai}\;{\bf n\cdot S_{\alpha}} 
+  \theta^{cs}_{I\alpha}\hat{n}_a({\bf \hat{n}}\times\nabla)_i\mu_{\alpha}
+ \kappa_{I\alpha} ({\bf \hat{n}}\times({\bf \hat{n}}\times\nabla))_a S_{\alpha}^i
+ \kappa_{I\alpha}'\delta_{ai}\nabla\cdot {\bf S_{\alpha\parallel}}\Big]
\label{J-I1}
\end{eqnarray}
These equations correspond to Eqs.~(8)-(10) of the main. Notice that for brevity in the main text we did not show the ``trivial'' contributions described by the first terms in Eqs.~\eqref{Sp}-\eqref{J-I1}, but kept only the relevant terms responcible for the spin-charge and spin-spin conversion.

\section{Spin-to-charge conversion: transport through the interface}

In this Section we present a solution of the boundary value problem for the spin-charge conversion in a metallic bilayer shown on Fig.~1 of the main text. As we explain in the main text, when a spin current $J_z^x(z)=-D\partial_zS^x(z)$, polarized along $x$-axis, flowing in $z$-direction crosses and interface at $z=0$ it generates a localized charge current in the y-direction via the IEE  $j_{yI}=\sigma^{sc}S^x(0)$, (see Eq. ($9$) of the main text). We want to describe the profile of the electrochemical potential and of the charge current in the case of a finite sample of width $W$ in the y-direction  (see Fig.~$1$ of the main text). To do so, we have to solve the Laplace equation with the BC
of vanishing  total current at the boundaries of the sample $y=\pm W/2$
\begin{eqnarray}
&&\nabla^2\mu=0\\
&&-\sigma^D\partial_y\mu(y,z)|_{y=\pm W/2}+j_{yI}\delta(z)=0.
\end{eqnarray}
By performing the Fourier transform with respect to the $z$-variable we obtain
\begin{eqnarray}
 \partial_y^2\mu(y,q)-q^2\mu(y,q)&=&0\\
\label{BC-vertical}
 -\sigma^D\partial_y\mu(y,q)|_{y=\pm W/2}+j_{yI}&=&0,
\end{eqnarray}
which has the following solution
\begin{equation}
 \mu(y,q)=C_1\cosh(qy)+C_2\sinh(qy),
\end{equation}
with the coefficients determined from the BC of Eq.~\eqref{BC-vertical}
\begin{equation}
 C_1=0\hspace{1cm}C_2=\frac{j_{yI}}{\sigma_D q\cosh(\frac{qW}{2})}.
\end{equation}
The solution for the electrochemical potential is given by the inverse Fourie transform,
\begin{equation}
\label{mu-vertical}
 \mu(y,z)=\int \frac{dq}{2\pi} e^{iqz}\frac{j_{yI}\sinh(qy)}{\sigma_D q\cosh(\frac{qW}{2})},
\end{equation}
which after performing the integral using the residue theorem can be written as
\begin{equation}
 \mu(y,z)=\frac{j_{yI}}{\sigma^D}\left[\Theta(z)\sum_{n=0}^{\infty}\frac{(-1)^ne^{-\frac{2(n\pi+\frac{\pi}{2})z}{W}}\sin(\frac{2(n\pi+\frac{\pi}{2})y}{W})}{2(n\pi+\frac{\pi}{2})}+
 \Theta(-z)\sum_{n=-\infty}^{-1}\frac{(-1)^ne^{-\frac{2(n\pi+\frac{\pi}{2})z}{W}}\sin(\frac{2(n\pi+\frac{\pi}{2})y}{W})}{2(n\pi+\frac{\pi}{2})}\right].
\end{equation}
After summing the series we find the following compact representation for the electrochemical potential
\begin{equation}
 \mu(y,z)=\frac{j_{yI}}{\sigma^D\pi}Im\left[\arctan(e^{\frac{\pi(-|z|+iy)}{W}})\right].
\end{equation}
The  charge current consists of two parts, the diffusive part in the bulk of the metals and the interfacial one,
\begin{eqnarray}
j_y&=&-\sigma^D\partial_y\mu(y,z)+j_{yI}(y,z)\\
j_z&=&-\sigma^D\partial_z\mu(y,z),
\end{eqnarray}
which can be written as follows.
\begin{eqnarray}
j_y&=&\frac{\sigma^D}{W}Re\left[\frac{ 1}{\cosh \left(\frac{\pi(z-iy)}{W}\right)}\right]j_{yI}+j_{yI}\delta(z)\\
j_z&=&\frac{\sigma^D}{W}Im\left[\frac{ 1}{\cosh \left(\frac{\pi(z-iy)}{W}\right)}\right]j_{yI}.
\end{eqnarray}
Finally, by using the integral representation \eqref{mu-vertical} we can calculate the total voltage drop across the sample:
\begin{equation}
\Delta V=\int\big[\mu(W/2,z)-\mu(-W/2,z)\big]dz=j_{yI}\frac{W}{\sigma^D}.
\end{equation}

\section{Spin-to-charge conversion: transport parallel to the interface}

Here we present a detailed analysis of our second example: the spin-charge conversion in a lateral hybrid structure made from a metallic film of thickness $W$, with part of its upper surface covered by an insulator with large SOC. 
This insulator generates an interface with ISOC on the top boundary at $z=W$  for $x>0$, while the rest ($x<0$) of the top boundary as well as the bottom boundary at $z=0$ remain ``trivial'' (see Fig.~$2$ in the main text). 
We assume a diffusive spin current polarized along $y$, flowing in the $x$-direction, that is, $J_x^y(x)=-D\partial_xS^y(x)$ with $S^y(x)=S_0e^{-x/l_s}$, where $l_s=\sqrt{D\tau_s}$ is the spin diffusion length of the normal metal. In order to calculate the induced electrochemical potential we have to solve the Laplace equation with two BC. One of them is the trivial condition of vanishing normal charge current at the bottom boundary, while in order to calculate the BC for the upper interface we have to combine the different BC for the diffusive and interfacial quantities (see Eqs.~(4), (9), and (11) of the main text). The Laplace equation with the BC read, 
\begin{eqnarray}
&&\nabla^2\mu=0\\
&&\sigma^D\partial_z\mu(x,z)|_{z=0}=0\\
&&\sigma^D\partial_z\mu(x,z)|_{z=W} = -D\partial_x\big[\theta^{sc}\Theta(x)S^y(x)\big],
\end{eqnarray}
By performing the Fourier transform with respect to the $x$-variable we obtain the following 1D problem
\begin{eqnarray}
 \partial_z^2\mu(q,z)-q^2\mu(q,z)&=&0\\
\sigma^D\partial_z\mu(q,z)|_{z=0}&=&0\\
\sigma^D\partial_z\mu(q,z)|_{z=W}&=& D\frac{\theta^{sc}S_0q^2}{(q-il_s^{-1})(q-i0^+)},
\end{eqnarray}
where infinitesimal imaginary shift $i0^+$ comes from the $\Theta(x)$ function which forces the pole at $q=0$ to be evaluated at the ``positive side'' of the real axis. The solution for the Fourier component of the electrochemical potential takes the following form
\begin{equation}
 \mu(q,z)=-\frac{\chi\theta^{sc}S_0q\cosh(qz)}{(q-il_s^{-1})(q-i0^+)\sinh(qW)},
\end{equation}
which in the real space reads,
\begin{equation}
\label{electrochemical}
 \mu(x,z)=\int \frac{dq}{2\pi} e^{iqx}\frac{\chi\theta^{sc}S_0q\cosh(qz)}{ (q-i0^+)(q-il_s^{-1})\sinh(qW)},
\end{equation}
After performing the integral using the residue theorem we find the following series representation for the electrochemical potential,
\begin{equation}
 \mu(x,z)=\chi\theta^{sc}S_0\left\{\Theta(x)\left[\frac{e^{-\frac{x}{l_s}}\cos(\frac{z}{l_s})}{\sin(\frac{W}{l_s})}+\sum_{n=0}^{\infty}\frac{(-1)^ne^{-\frac{n\pi x}{W}}\cos(\frac{n\pi z}{W})}{n\pi-\frac{W}{l_s}}\right]+
 \Theta(-z)\sum_{n=-\infty}^{-1}\frac{(-1)^ne^{-\frac{n\pi x}{W}}\cos(\frac{n\pi z}{W})}{n\pi-\frac{W}{l_s}}\right\}.
\end{equation}
This series can be expressed in terms of the the Lerch transcendent function \cite{Erdelyi},
\begin{equation}
 \Phi(z,s,a)=\sum_{n=0}^{\infty}\frac{z^n}{(n+a)^s},
\end{equation}
as follows
\begin{eqnarray}\nonumber
 \mu(x,z)&=&\chi\theta^{sc}S_0\left\{\Theta(x)\left[\frac{e^{-\frac{x}{l_s}}\cos(\frac{z}{l_s})}{\sin(\frac{W}{l_s})}+\frac{1}{2\pi}\left\{\Phi\left(-e^{-\frac{\pi(x+iz)}{W}},1,-\frac{W}{l_s\pi}\right)
 +\Phi\left(-e^{\frac{\pi(-x+iz)}{W}},1,-\frac{W}{l_s\pi}\right)\right\}\right]\right.\\
\label{mu-lateral}
 &&\left.  - \frac{1}{2\pi}\Theta(-z)\left\{e^{\frac{\pi(x+iz)}{W}}\Phi\left(-e^{\frac{\pi(x+iz)}{W}},1,-\frac{W}{l_s\pi}\right)
 +e^{\frac{\pi(x-iz)}{W}}\Phi\left(-e^{-\frac{\pi (x-iz)}{W}},1,-\frac{W}{l_s\pi}\right)\right\}\right\},
\end{eqnarray}
which is our final results for the electrochemical potential in the lateral structure.
\begin{figure}
\begin{center}
\includegraphics[width=3.2in]{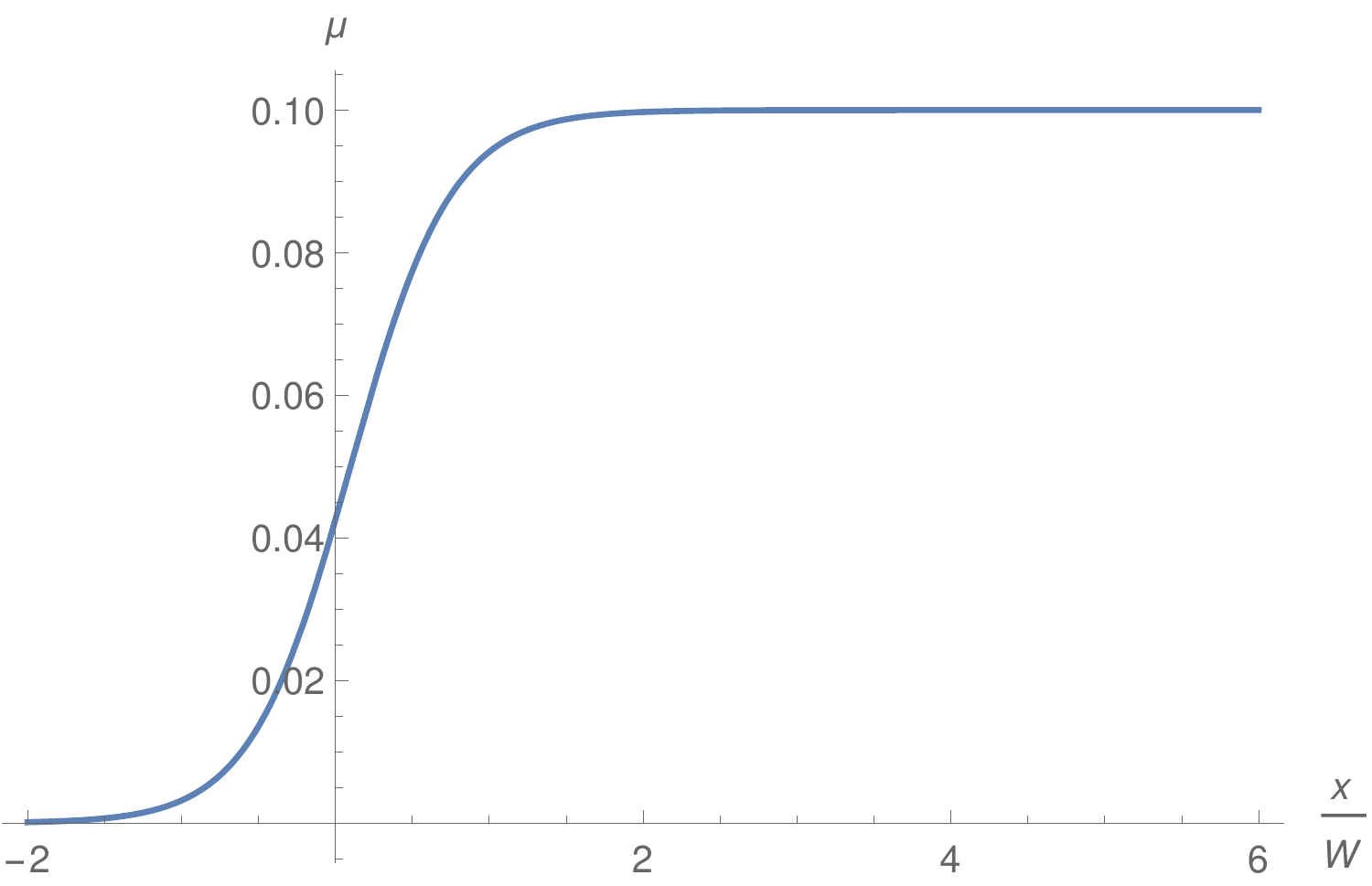}
\caption{Profile of the potential $\mu(x,0)$ at $z=0$ as a function of x  for $l_s=0.1W$.}
\label{Voltage}
\end{center}
\end{figure}

In Fig.~\ref{Voltage} we show the potential $\mu(x,0)$ at $z=0$ as a function of $x$. We observe that the longitudinal voltage drop is formed of the scale of the order of the sample width $W$. To find the value of this voltage drop explicitly we the asymptotic behavior of Eq.~(\ref{electrochemical}) for $|x|>>W$. In this limit the electrochemical potential can be written as follows
\begin{equation}
  \mu(x,z)\approx\int \frac{dq}{2\pi} e^{iqx}\frac{\chi\theta^{sc}S_0}{ (q-i0^+)il_s^{-1}W}=\chi\theta^{sc}S_0\frac{l_s}{W}\Theta(x),
\end{equation}
so the total voltage drop reads,
\begin{equation}\label{DeltaV}
 \Delta V=\mu(\infty,z)-\mu(-\infty,z)=\chi\theta^{sc}S_0\frac{l_s}{W}.
\end{equation}

\subsection{Lateral structure with ISOC-gate of a finite length}
Finally, we analyze the same problem for the case when the ``ISOC-gate'' has a finite length $L$. If the ISOC acts within a finite region $0<x<L$ on the top surface the corresponding BC is modified as follows
\begin{equation}
 \sigma^D\partial_z\mu^{L}|_{z=W} = -D\partial_x\big[\theta^{sc}[\Theta(x)-\Theta(x-L)]S^y(x)\big].
\end{equation}
In this case the final solution for the electrochemical potential takes the form
\begin{equation}
 \mu^L(x,z)=\mu^\infty(x,z)-e^{-\frac{L}{l_s}}\mu^\infty(x-L,z),
\end{equation}
where $\mu^\infty(x,z)$ is the potential for $L\to\infty$, given by Eq.~\eqref{mu-lateral}.
\begin{figure}
\begin{center}
\includegraphics[width=3.2in]{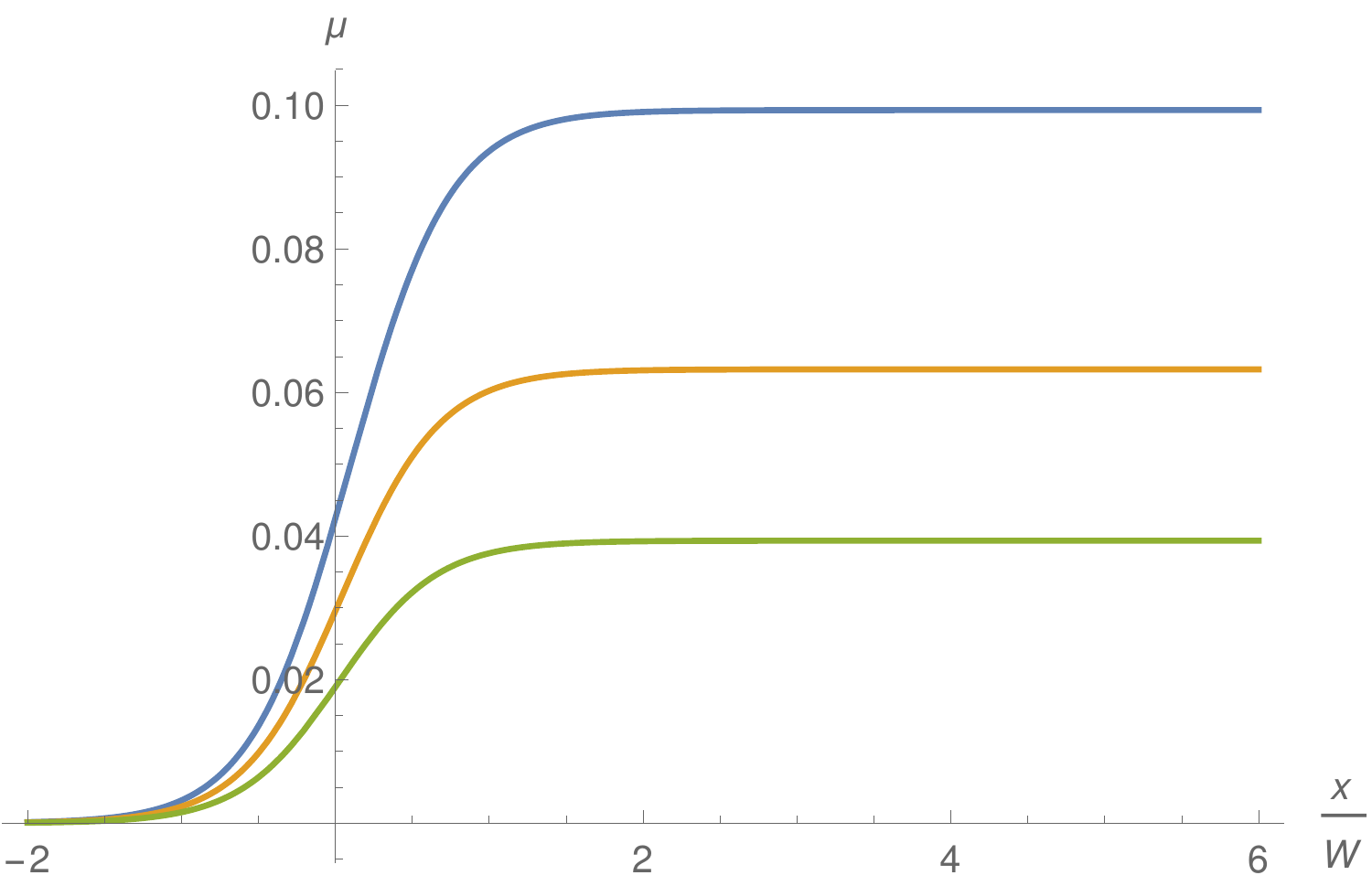}
\caption{Profile of the potential $\mu^L(x,0)$ at $z=0$ as a function of x  for $l_s=0.1W$ for different values of $L$, $L=5l_s$ (blue), $L=l_s$ (yellow) and $L=0.5l_s$ (green).}
\label{VoltageL}
\end{center}
\end{figure}
In Fig.~\ref{VoltageL} we show the potential $\mu^L(x,0)$ at $z=0$ as a function of $x$  for $l_s=0.1W$ for different values of $L$. For the corresponding lateral voltage drop we find,
\begin{equation}
 \Delta V^{L}=\Delta V^\infty(1-e^{-\frac{L}{l_s}}),
\end{equation}
where $\Delta V^\infty$ is defined after Eq.~\eqref{DeltaV}.

\begin{figure}
\begin{center}
\includegraphics[width=3.5in]{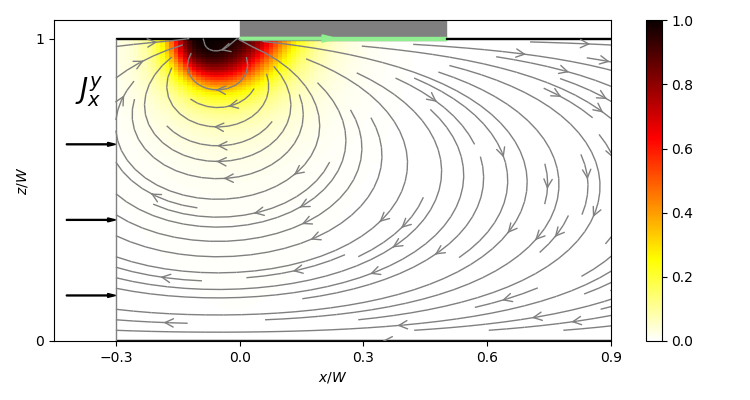}
\caption{Charge flow generated by the spin current $J_x^y(x)$ in the metallic film with a insulator of length $L$ (shown in gray) deposited on its top surface. The density plot shows the current strength.}
\label{current}
\end{center}
\end{figure}

Finally in Fig.~\ref{current} we show the profile of the charge flow generated by the spin current $J_x^y(x)$ in the metallic film with an insulator of length $L$,  where we have chosen $L=l_s=0.5W$.


\begin{thebibliography}{57}
\bibitem [{BCn()}]{BCnotation-note}%
  \BibitemOpen
  \href@noop {} {}\bibinfo {note} {All quantities entering the BC and labeled
  by the material index $\alpha=1$ and 2 are evaluated at a given point of the
  interface as the right and left limits, respectively, i.~e. $\hat{\bf
  n}\cdot\rv=\pm 0$. Throughout this paper for brevity we omit the spatial
  arguments in the BC.}\BibitemShut {Stop}%
\bibitem [{n-d()}]{n-derivative-note}%
  \BibitemOpen
  \href@noop {} {}\bibinfo {note} {The allowed by symmetry terms involving
  normal derivatives can always be eliminated using the BC of
  Eqs.~\eqref{BC-charge}-\eqref{BC-spin-p}, and therefore they do not appear in
  the expression for the localized observables.}\BibitemShut {Stop}%
\bibitem [{pro()}]{projector-note}%
  \BibitemOpen
  \href@noop {} {}\bibinfo {note} {More formally this condition is expressed in
  terms of the interface projection $J_{iI}^a\mapsto (\delta_{ik} -
  \hat{n}_i\hat{n}_k)J_{kI}^a$, which is implicitly assumed in
  Eq.~\eqref{J-I}.}\BibitemShut {Stop}%
\bibitem [{sig()}]{sigma-theta-note}%
  \BibitemOpen
  \href@noop {} {}\bibinfo {note} {To arrive at this form of BC we have used
  the relation $\sigma^{sc}=D\theta^{sc}$ which follows from
  Eq.~\eqref{theta-sigma} for the case of a metal-insulator
  interface.}\BibitemShut {Stop}%
\end{thebibliography}

\end{document}